\documentclass[sigconf]{acmart}

\usepackage{booktabs} 

\usepackage{graphicx,color,colortbl,psfrag,subfig}
\usepackage{multirow}
\usepackage{algorithm}
\usepackage{algpseudocode}

\usepackage{amssymb}
\usepackage{amsmath}
\usepackage{amsfonts}

\DeclareMathOperator*{\argmax}{arg\,max}

\usepackage{arydshln}

\copyrightyear{2017} 
\acmYear{2017} 
\setcopyright{acmcopyright}
\acmConference{ADKDD'17}{August 14, 2017}{Halifax, NS, Canada}\acmPrice{15.00}\acmDOI{10.1145/3124749.3124757}
\acmISBN{978-1-4503-5194-2/17/08}

\begin{document}
\title[]{MM2RTB: Bringing Multimedia Metrics to Real-Time Bidding}

\author{Xiang Chen}
\affiliation{\institution{National University of Singapore}}
\email{chxiang@comp.nus.edu.sg}

\author{Bowei Chen}
\affiliation{\institution{University of Lincoln}}
\email{bchen@lincoln.ac.uk}

\author{Mohan Kankanhalli}
\affiliation{\institution{National University of Singapore}}
\email{mohan@comp.nus.edu.sg}

\begin{abstract}

In display advertising, users' online ad experiences are important for the advertising effectiveness. However, users have not been well accommodated in real-time bidding (RTB). This further influences their site visits and perception of the displayed banner ads. In this paper, we propose a novel computational framework which brings multimedia metrics, like the \emph{contextual relevance}, the \emph{visual saliency} and the \emph{ad memorability} into RTB to improve the users' ad experiences as well as maintain the benefits of the publisher and the advertiser. We aim at developing a vigorous ecosystem by optimizing the trade-offs among all stakeholders. The framework considers the scenario of a webpage with multiple ad slots. Our experimental results show that the benefits of the advertiser and the user can be significantly improved if the publisher would slightly sacrifice his short-term revenue. The improved benefits will increase the advertising requests (demand) and the site visits (supply), which can further boost the publisher's revenue in the long run.    

\end{abstract}

%
%

%



\keywords{Display advertising; real-time bidding; multimedia metrics}

\maketitle

\section{Introduction}
\label{sec:introduction}

The fundamental issue of computational advertising is to select a proper ad from a set of ad candidates. Introduced in 2009, RTB has become the norm of selling online users' page views (also called \emph{impressions}) in display advertising, where the ad with the highest bid wins the auction for the impression. Obviously, the existing RTB system is biased towards the publisher. However, the effectiveness of displayed ads remains debatable, in terms of the benefits of the advertiser and the user. According to~hubspot~\cite{hubspot_annoy_ad}, a recent survey shows only 2.8\% of participants think banner ads are relevant to what they read online and another 33\% even think many ads completely intolerable. In the year 2015, ad blocking has grown by 41\% globally. On one hand, the user gets annoyed by the improper ads. On the other hand, the advertiser gets harmed by ineffective ad delivery. Moreover, for those webpages with multiple ad-slots, the ad slots are sold separately through individual auctions~\cite{Ben-Zwi_2015}. As a result, two ads about the same product but belong to different advertisers are likely to be displayed within the same webpage. The competitive ads have a negative effect on user's brand conception~\cite{burke1988competitive}. 


Motivated by the above observations, we argue that the context matters in display advertising. Literature in marketing and consumer psychology has already shown that the contextual relevance between the content of host webpage and the ads makes a large difference in their clickability~\cite{chatterjee2003modeling}, and it also has a leading effect on the user's online experience~\cite{mccoy2007effects}. Through a large-scale live test, combining contextual relevance and targeting strategies is able to increase the ad CTR~\cite{lu2016combining}. To increase user's engagement towards the displayed ads, we introduce the \emph{contextual relevance}. Through a series of eye-tracking experiments, recent research found that users tend to avoid ads in web search and surfing~\cite{broder2008search} and they intentionally avoid looking at such ads even when they are designed to be attention-grabbing~\cite{sajjacholapunt2014influence}. This is also known as~\emph{ad overlook}. Moreover, Owen et al.~\cite{owens2011text} explored the relationship between ads location and the degree of blindness -- the phenomenon of website users actively ignore web banner ads -- and found that users tend to ignore ads located on the bottom and right area. Intuitively, any ad that fails to capture the user's attention will be ineffective in delivering information. To ensure that the user will notice the displayed ad, we introduce the \emph{visual saliency}. Image memorability has been shown to be a stable and intrinsic property of images that is shared across different viewers~\cite{khosla2012image}. Thus, the memorable ads will be easily recalled by the user. To enhance the user's brand conception towards the displayed ad, we introduced the \emph{ad memorability}.

\begin{figure}[t]
\centering
\includegraphics[width=0.95\linewidth]{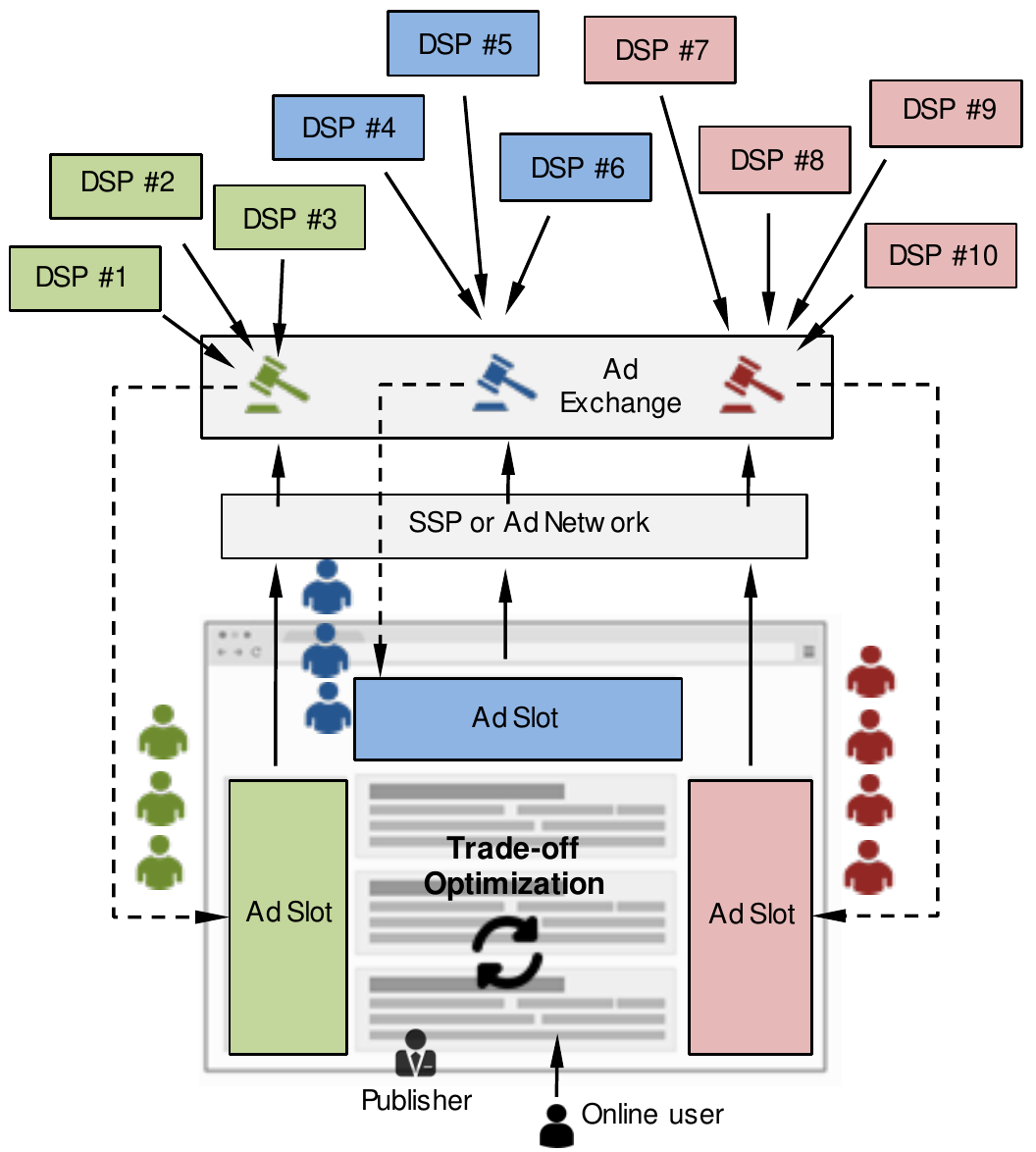}
\vspace{-5pt}
\caption{Schematic view of trade-offs optimization for RTB.}
\label{fig:optimization_framework}
\vspace{-15pt}
\end{figure}

Fig.~\ref{fig:optimization_framework} illustrates how multimedia metrics are incorporated into RTB. When an online user visits a publisher's website, the publisher's web server loads up a webpage and sends an HTML code to the user's browser so that the latter will know where to get the content and how to format it. In the meantime, the publisher's webpage information, such as the publisher ID, the site ID, and the dimensions of ad slots, together with the user's cookie ID, will be passed to and be accessed by an RTB-enabled supply-side platform (SSP) or ad network. Each ad slot is considered to generate an impression and it will be treated independently in the selling. For an ad slot, an auction is started within ad exchange by requesting bids from demand sources such as the demand-side platforms (DSPs) or individual advertisers, the winning advertiser will be able to have his ad displayed to the user.\footnote{Individual advertisers typically do not directly participate in advertising auctions but entrust some demand-side companies to bid on their behalf.} The whole RTB process includes user identification, auction and ad display, usually be finished in 10 to 100 milliseconds~\cite{YYuan_2014}. The displayed ads on a single webpage come from individual auctions and the selection criteria is mainly based on bids. The visual effects, particularly, their interactive effects are not well considered. This will affect the user's experience as well as the advertising effectiveness for each advertiser, which further affect the publisher's long-term revenue. Therefore, we propose a trade-offs optimization framework in this paper. 


In this study, we focus on improving the effectiveness of display advertising with considering all stakeholders' benefits. We consider multimedia metrics to measure some stakeholders' benefits, such as the ad memorability, the contextual relevance, and the visual saliency. To the best of our knowledge, this is  one of the few studies that combines multimedia techniques and auction theory together. Different from our previous work~\cite{chen2017RTBoptimizing}, which focuses on single-slot display advertising, we extend the idea under multi-slot scenario. 


\section{Preliminaries}
\label{sec:model}

The key concepts and notations are briefly explained as below:

\textbf{\emph{Publisher}} -- a company or individual who owns a webpage which has slots to host online ads. A webpage is denoted by $w$.  

\textbf{\emph{Ad slot}} -- a rectangular area within a webpage where a banner ad can be displayed to online users. In this paper, we consider the case that a webpage contains multiple ad slots. A slot on webpage $w$ is denoted by $s \in S_w$, where $S_w$ is the set of slots.  

\textbf{\emph{Advertiser}} -- a company or individual who wants to display his banner ad on the publisher's webpage. The notation $a_{s,w,l}$ represents advertiser $l$ bids for slot $s$ on webpage $w$. 

\textbf{\emph{Company}} -- a company is defined based on the URL domain of the ad landing page. If two different ads point to a same URL domain, they are recognized as one company's ads. 

\textbf{\emph{Topic}} -- a cluster in which ads are about similar products or services. We use the ad landing page texts to represent the ad information, and employ the term frequency-inverse document frequency (TF-IDF) and K-Means methods~\cite{MacKay2003} to cluster all ads into 24 topics\footnote{According to the IAB content taxonomy~\cite{IAB_content_taxonomy}, online contents can be broadly grouped into 24 topics.}. 

\textbf{\emph{Competitive advertiser}} -- an advertiser is said to be a competitive advertiser to the other if both advertisers have at least one ad being clustered into a same topic group. For example, if iPhone 7 and Galaxy S8 are clustered into the smartphone topic, Apple and Samsung are competitors. However, even though iPhone 6S and iPhone 7 join RTB as different bidders, they are not competitive advertisers because both come from Apple.

\section{Trade-offs Optimization}
\label{sec:trade_offs_optimization}

Suppose that a webpage $w$ has a set of slots $\mathcal{S}_w$ to host banner ads. When an online user visits it, $|\mathcal{S}_w|$ impressions are created and are auctioned off separately. The model optimizes the trade-offs among multiple stakeholders over multiple slots by the following steps:

\textbf{\emph{Step 1}} --  Create a matrix $\boldsymbol{\Omega}$ that contains all advertisers who join RTB campaigns for impressions from any slot(s) of webpage $w$. As shown in Fig.~\ref{fig:optimization_framework}, the example webpage contains three ad slots. For this online user's visit, there are: three advertisers bid for the first slot (in blue color); three advertisers bid for the second slot (in green color); and three advertisers bid for the third slot (in red color). Therefore, $\boldsymbol{\Omega}$ can be expressed as
\[
\boldsymbol\Omega =
\left(
\begin{array}{c:c:c}
a_{1,w,1} & a_{2,w,1} & a_{3,w,1} \\
a_{1,w,1} & a_{2,w,1} & a_{3,w,2} \\
a_{1,w,1} & a_{2,w,1} & a_{3,w,3} \\
a_{1,w,1} & a_{2,w,1} & a_{3,w,4} \\
a_{1,w,1} & a_{2,w,2} & a_{3,w,1} \\
a_{1,w,1} & a_{2,w,2} & a_{3,w,2} \\
		& \cdots  &		   \\
a_{1,w,3} & a_{2,w,3} & a_{3,w,4} \\		
\end{array}
\right),
\]
where $a_{1,w,3}$ represents advertiser 3 who bids for the first slot on webpage $w$. The column of $\boldsymbol\Omega$ represents a specific slot and its row represents a specific combination of candidate advertisers, so the size of $\boldsymbol\Omega$ is $z \times |\mathcal{S}_w|$, where $z = \prod_{s \in \mathcal{S}_w} \binom 1{|\mathcal{A}_{s,w}|}$, and $\mathcal{A}_{s,w}$ is the set of advertisers who join the RTB campaign for slot $s \in \mathcal{S}_w$.

\textbf{\emph{Step 2}} --  Create a subset matrix $\boldsymbol{A}$ from $\boldsymbol{\Omega}$ by removing the rows which contain competitive advertisers. The size of $\boldsymbol{A}$ is $q \times |\mathcal{S}_w|$, where $0 \leq q \leq z$. 

\textbf{\emph{Step 3}} --  Compute the rank score for each combination of candidate advertisers in $\boldsymbol{A}$. For $i = 1, \cdots, q$, the rank score $r_i$ is
\begin{equation}
r_i =  \boldsymbol{\gamma}^{*\top} \boldsymbol{x}_i,
\end{equation} 
where $\boldsymbol{x}_i$ is the vector of calculated values of metric variables for candidate advertisers on the $i$th row of $\boldsymbol{A}$, and $\boldsymbol{\gamma}^{*}$ is the vector of their optimal weights. More details about $\boldsymbol{x}_i$ and $\boldsymbol{\gamma}^*$ are discussed in Sections~\ref{sec:calc_metric}-\ref{sec:optimal_weights}, respectively.

\textbf{\emph{Step 4}} --  The optimal selection of advertisers are $\boldsymbol{A}(i^*,:)$, where
\begin{equation}
i^* = \argmax_i \{r_i\}_{i=1}^{q}.
\end{equation}

\section{Calculation of Metric Variables}
\label{sec:calc_metric}

Six metric variables are considered in the trade-offs optimization, including the publisher's revenue, the advertisers' utility, the ads' memorability, the CTR, the contextual relevance, and the ads' saliency. Different to~\cite{chen2017RTBoptimizing}, as multiple slots are considered in this study, each variable's value is the sum of the corresponding metric values of selected advertisers, and each slot has the same weights in the variable value calculation. Algorithm~\ref{algo:metric_calc} illustrates the calculation of metric variables for a given vector of candidate advertisers $\boldsymbol{A}(i,:)$. Note that, all the six metrics are normalized before proceeded to the objective function. 

As discussed earlier, impressions are auctioned off separately. If $a \in \boldsymbol{A}(i,:)$ and she bids for slot $s$, we can employ the method discussed in~\cite{chen2017RTBoptimizing} by creating several pseudo slots and then calculate his (potential) payment by a GSP auction model. An advertiser's short-term benefit can be measured by his utility, which is defined as the difference between his value and payment. An advertiser's long-term benefit can be measured by the ad's memorability~\cite{Neumeier_2005} -- it shows how likely the user will remember the advertiser's ad in a few weeks or months. Same as~\cite{chen2017RTBoptimizing}, we employ the MemNet~\cite{khosla2015understanding} model to predict the visual memorability of the ad image. 
The rest three metrics represent the user's benefit. The ad CTR is defined as the number of clicks on the advertiser's ad divided by the number of displays, whose value is usually given by data or can be estimated from historical advertising records. It is an ad quality metric -- a high CTR means that the advertiser's ad is attractive or more relevant to the user's needs. The contextual relevance measures if the ad content is more or less relevant to its hosting webpage content. Note that, the textual information of the webpage is represented by the combination of webpage title, keywords, description and main content. Different to~\cite{chen2017RTBoptimizing}, we use the TF-IDF~\cite{MacKay2003} here to measure the similarity of textual contents between the ad and the webpage. This is because ~\cite{chen2017RTBoptimizing} uses webpage title, keywords and description to construct the textual information of the webpage, where the adopted method Takelab~\cite{vsaric2012takelab} is good at measuring the similarity between short text snippets. The ad saliency metric measures whether the ad image can be easily spotted within its hosting webpage. We use the minimum barrier salient (MBS) object detection method~\cite{zhang2015minimum} to calculate the saliency of the ad image. 
For each pair of webpage and ad candidate, we embed the ad into the webpage and use the MBS method to calculate the saliency map. We take the mean value of each pixel within the ad area and view this value as the saliency score of the ad candidate.

\section{Determination of Optimal Weights}
\label{sec:optimal_weights}

Suppose that $n$ site visits have been observed, the optimal weights $\boldsymbol{\gamma}^*$ can be obtained by solving the following optimization problem:   
\begin{align}
\boldsymbol\gamma^* = \argmax_{\boldsymbol\gamma} 
& \ 
\sum_{j=1}^{n} \boldsymbol{\gamma}^{\top} \boldsymbol{x}_{i^*}^{\{j\}}, \\
\text{s.t.} 
& \ 0 \leq \gamma_k \leq 1, k = 1, \cdots, 6,  \label{eq:weight_nonnegative}\\
& \ \boldsymbol{\gamma}^{\top} \boldsymbol{1} = 1,  \label{eq:weight_unity}\\
& \ |\xi_{1}| \leq |\theta_{1}|, \theta_1 \leq 0, \label{eq:threshold_revenue} \\
& \ \xi_{k} \geq \theta_{k}, \theta_k \geq 0, k = 2, \cdots, 6, \label{eq:threshold_others} 
\end{align}
where $\xi_k$ is defined by
\begin{equation}
\label{eq:variable_change_defination}
\xi_k = \frac{\sum_{j=1}^{n} 
\Big( \boldsymbol{x}_{i^*}^{\{j\}}(k) - \boldsymbol{x}_{i^\neg}^{\{j\}}(k)
\Big)
}{
\sum_{j=1}^{n}
\boldsymbol{x}_{i^\neg}^{\{j\}}(k)
}, \ \ \ k = 1, \cdots, 6.
\end{equation}

The terms $i^*$ and $i^\neg$ are the indexes of optimal solution and ground truth (i.e., the selected ads using the existing RTB system) for candidate matrix $\boldsymbol{A}^{\{j\}}$ for the $j$th observation in the training data, and $\theta_k$ is the threshold value for specifying the bound of changes in variable $k$. The optimal weights maximize the sum of rank scores of the select advertisers from all auctions in the training set. Eqs.~(\ref{eq:weight_nonnegative})-(\ref{eq:weight_unity}) ensure each variable has an impact and its impact has an upper bound. Eqs.~(\ref{eq:threshold_revenue})-(\ref{eq:threshold_others}) further specify the maximum decrease of the publisher's revenue and the minimum increases for other variables.

\begin{algorithm}[t]
\footnotesize
\caption{Calculation of metric variables for a given $\boldsymbol{A}(i,:)$.}
\label{algo:metric_calc}
\begin{algorithmic}[1]
\State $\boldsymbol{x}_i(1) \leftarrow \sum_{a \in  \boldsymbol{A}(i,:)} \textrm{Payment}_a$ \Comment{GSP auction~\cite{chen2017RTBoptimizing}}
\State $\boldsymbol{x}_i(2) \leftarrow \sum_{a \in  \boldsymbol{A}(i,:)} (\textrm{Value}_a - \textrm{Payment}_a)$
\State $\boldsymbol{x}_i(3) \leftarrow \sum_{a \in  \boldsymbol{A}(i,:)} \textrm{MemScore}_a$  \Comment{MemNet~\cite{khosla2015understanding}}
\State $\boldsymbol{x}_i(4) \leftarrow \sum_{a \in  \boldsymbol{A}(i,:)} \textrm{CTR}_a$
\State $\boldsymbol{x}_i(5) \leftarrow \sum_{a \in  \boldsymbol{A}(i,:)} \textrm{RevScore}_a$  \Comment{TF-IDF~\cite{MacKay2003}}
\State $\boldsymbol{x}_i(6) \leftarrow \sum_{a \in  \boldsymbol{A}(i,:)} \textrm{SaliencyScore}_a$ \Comment{MBS~\cite{zhang2015minimum}}
\State \Return $\boldsymbol{x}_i$ 
\end{algorithmic}
\end{algorithm}

\begin{table}[t]
\footnotesize
\centering
\caption{Multimedia datasets. The crawler type (I) collects as many webpages as possible from a seed URL and the crawler type (II) repeatedly accesses to a set of particular webpages. }
\label{tab:multimedia_datasets}
\vspace{-7pt}
\begin{tabular}{r|r|r|r|r|r}
\hline
\multicolumn{2}{r|}{Website} & Yahoo & Yahoo & MSN & MSN \\
\hline
\multicolumn{2}{r|}{Crawler type} & I & II & I & II \\
\multicolumn{2}{r|}{From} & \hspace*{-2pt} 20 Jan 2017 & \hspace*{-2pt} 20 Jan 2017 & \hspace*{-2pt}  20 Jan 2017 & \hspace*{-2pt}  20 Jan 2017\\
\multicolumn{2}{r|}{To}   & \hspace*{-2pt} 30 Jan 2017 & \hspace*{-2pt} 30 Jan 2017 & \hspace*{-2pt} 30 Jan 2017 & \hspace*{-2pt} 30 Jan 2017\\
\multicolumn{2}{r|}{Location} & Singapore & Singapore & Singapore & Singapore\\
\multicolumn{2}{r|}{\# of webpages} & & & &\\ 
\multicolumn{1}{r}{}  & with 1 slot  & 1,481 & 1,909 & 798 & 686\\
\multicolumn{1}{r}{}  & with 2 slots & 1,978 & 3510 & 1,519 & 3,689\\
\multicolumn{1}{r}{}  & with 3 slots & 1,173 & 4329 & 3,633 & 146\\
\multicolumn{1}{r}{}  & with $\geq 4$ slots & 1,599 & 2,468 & 41 & 241\\
\multicolumn{2}{r|}{\# of total impressions} & 15,836 & 31,951 & 14,899 & 9,466\\
\multicolumn{2}{r|}{\# of unique advertisers} & 692 & 631 & 160 & 163\\
\multicolumn{2}{r|}{\# of total companies} & 475 & 431 & 96 & 99\\
\hline
\end{tabular}
\vspace{-7pt}
\end{table}

\section{Experiments}
\label{sec:experiments}
This section presents our datasets, empirical findings, experimental settings, 
and overall results of trade-offs optimization.

\subsection{Datasets}
\label{sec:data_description}
We use two types of datasets to simulate a real-world advertising system: the multimedia dataset represents the interaction between the publisher and the user, and the auction dataset shows the interplay between the publisher and the advertiser.  

We have collected data from Yahoo and MSN over the period from 20 January to 30 January in 2017 to construct our multimedia datasets. All the multimedia datasets were collected in Singapore. For each website, we designed two data crawlers, crawler type (I) started from a given seed URL, and used breadth-first-search to collect as many different webpages as possible. To ensure the diversity of webpages and banner ads, we started from the home page of each website, which contains multiple categories of contents.\footnote{
\small
\begin{tabular}{ll}
Yahoo: &\href{https://sg.yahoo.com}{\tt\small https://sg.yahoo.com}\\
MSN:&\href{http://www.msn.com/en-sg/}{\tt\small http://www.msn.com/en-sg/}
\end{tabular}
}
Crawler type (II) repeatedly accessed to a set of particular webpages at a frequency of every 5 minutes. In our experiment, this webpage set was made up of the homepage of the sub categories, such as Yahoo news, Yahoo finance, Yahoo sports and so on, which were the most frequently viewed webpages by the users. These two types of data crawler were designed for different purpose: crawler type (I) indicated how the ad changed with different webpage contents, while crawler type (II) represented how the ad changed with different timestamps. Since the ads on both websites are dynamically embedded, we used Python, Selenium and Chromedriver to collect the webpage and ads that displayed to the user as shown in real-world. Note that, in collecting the data, the browser was set in the privacy mode, which disabled browsing history, web cache and data storage in cookies so that the collected banner ads were not affected by the previous page views. In each dataset, we extracted the ads from their webpages to create a set of banner ads and a set of webpages with blank ad slots. For each webpage, the collected data includes title, keywords, description, whole webpage snapshot, ad image. We also crawled title, keywords and description from the ad landing page (i.e., the delivered webpage if an ad is clicked by user). Note that we did not consider animation ads in our banner ads. According to the definition in section~\ref{sec:model}, we were able to figure out the advertiser set, company set and competitive advertiser-pair set. Our multimedia datasets have been released publicly and more information can be found at: ~\href{https://github.com/boweichen/MultiSlotRTBMultimediaDataset}{\small\tt https://github.com/boweichen/MultiSlotRTBMultimediaDataset}.

\begin{figure}[t]
\centering
\includegraphics[width=0.975\linewidth]{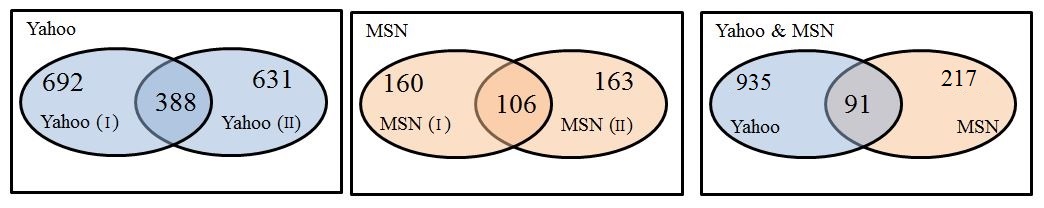}
\vspace{-7pt}
\caption{The intersection of ad sets within each website and across different websites}
\vspace{-7pt}
\label{fig:ad_set_union}
\end{figure}

\begin{table}[t]
\footnotesize
\centering
\caption{Statistic about webpage with different scenarios}
\vspace{-7pt}
\label{tab:webpage_statistic}
\begin{tabular}{r|r|r|r|r}
\hline
    Ad dataset & Yahoo I  & Yahoo II & MSN I & MSN II\\
    \hline
    \# of total webpages & 4,750 & 10,307 & 5,193 & 4,076 \\
    \# of webpages with scenario 1) & 786 & 2,169 & 3,669 & 1,480\\
    \# of webpages with scenario 2) & 155 & 924 & 572 & 238\\
    \# of webpages with scenario 3) & 412 & 836 & 158 & 82\\
\hline
\end{tabular}
\end{table}

\begin{figure}[t]
\centering
\vspace{-7pt}
\includegraphics[width=0.975\linewidth]{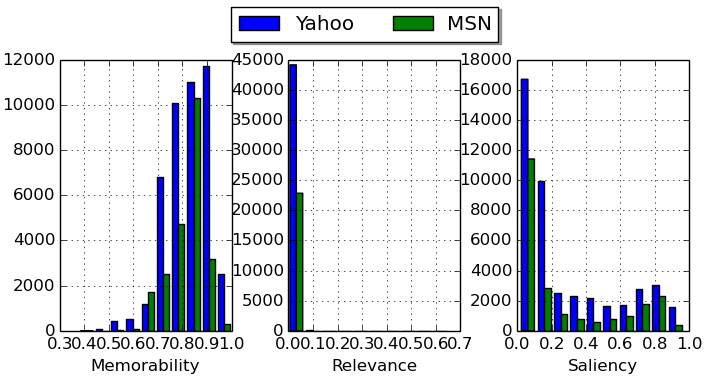}
\vspace{-7pt}
\caption{The distribution of multimedia metrics on the two websites, where X-axis represents the value of variables, and Y-axis represents the number of webpages-ad pairs.}
\label{fig:variable_distribution}
\vspace{-15pt}
\end{figure}

\begin{figure}[t]
\centering
\includegraphics[width=0.975\linewidth]{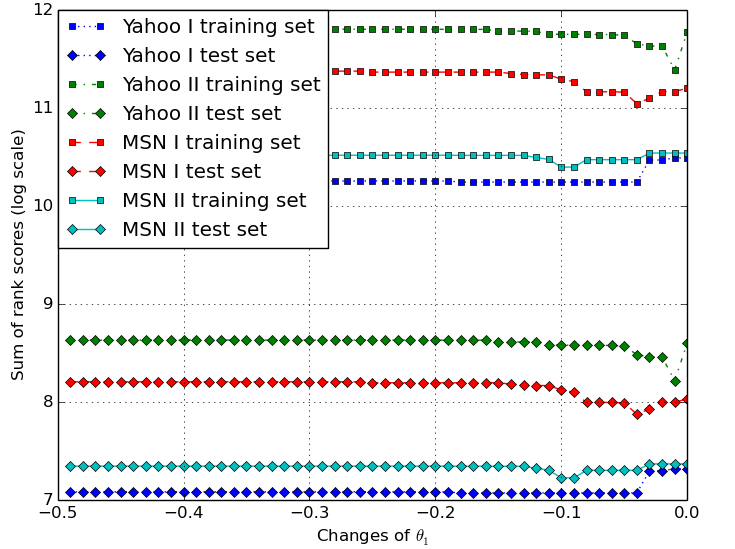}
\vspace{-7pt}
\caption{The effect of $\theta_{1}$ on the sum of total rank scores of the selected advertisers in the auctions. Note that we only show $\theta_{1}$ from 0.0 to -0.50, since the sum of rank score will not change when $\theta_{1}$ is greater than a particular value.}
\vspace{-10pt}
\label{fig:sum_rank_score_with_theta}
\end{figure}

\begin{figure*}[t]
\centering
\includegraphics[width=0.975\linewidth]{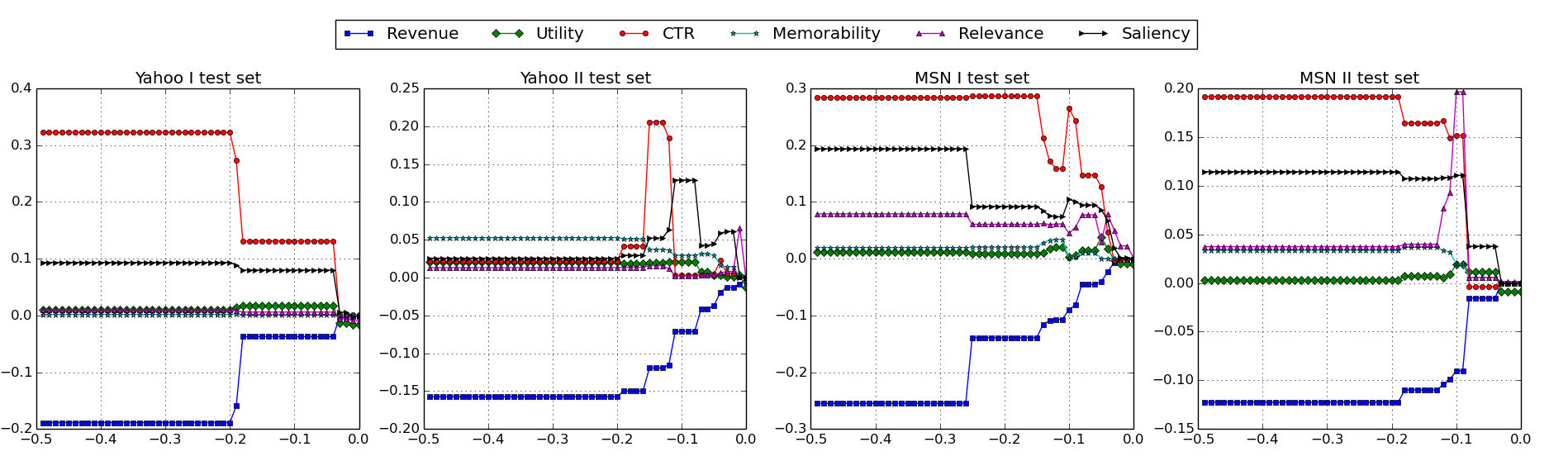}
\vspace{-7pt}
\caption{The effect of $\theta_{1}$ on the changes of variables, where X-axis represents the change of $\theta_{1}$ and Y-axis represents the change of corresponding variables. Note that: 1) we only show $\theta_{1}$ from 0.0 to -0.50, since the variables will not change when $\theta_{1}$ is greater than a particular value; 2) we do not show the changes of variables in the training set since the results of the training sets are similar to that of the test sets.}
\vspace{-7pt}
\label{fig:variable_change_with_theta}
\end{figure*}

\subsection{Empirical Findings}

Table~\ref{tab:multimedia_datasets} describes the data pre-processing results on the four multimedia datasets. As we can see, the total number of ad impressions of each website under each crawler type is 15,836, 31,951, 14,899, 9,466, while the number of unique advertisers is 692, 631, 160, 163, respectively. Although there are a large number of impressions in the ad network, only a few ads are displayed. We find that some ads re-appear from time to time. The repetitive display strategy reinforces users' memory for branding but it is also a source of intrusiveness into users' online experience~\cite{campbell2008shut}. Another interesting finding is illustrated in Fig.~\ref{fig:ad_set_union}. As can be seen, there are 388 advertisers' ads appearing in both Yahoo datasets out of all the 935 advertisers, and 91 out of 217 in two MSN datasets. The intersection within each website indicates that: the content of the host webpages affects the displayed ads and same ads only appeared in the specific webpages. And also, there are 91 advertisers' ads appearing on both Yahoo and MSN, which shows the existence of interactions between different DSPs and websites. 

Since user's memory towards displayed ads varies in various contexts, the repetitive ads can enhance the brand perception while competitive ads will result in the opposite effect~\cite{burke1988competitive}. Given that each ad slot is auctioned separately, the following three scenarios are possible in a multi-slot webpage: 1) two ads with the same landing page are displayed; b) two ads with different landing page but belonging to the same company are displayed; 3) two competitive ads are displayed. Note that there exists overlaps among the above three scenarios. For example, there are four ads ${ad_{1}, ad_{2}, ad_{3}, ad_{4}}$ displayed in the webpage, where $ad_{1}, ad_{2}$ are promotions to buy the latest Apple Iphone 7 in Apple store, $ad_{3}$ is a promotion to buy a Iphone 6, while $ad_{4}$ is about how to buy Samsong Galaxy S7. All the above three scenarios occur. Table.~\ref{tab:webpage_statistic} summarizes the statistic of the three scenarios in our four multimedia datasets. When comparing the number of webpages in scenario 1) and scenario 2), we can find that: when two ads from a same company are displayed within the same webpage, the probability that they are the same ad is higher than that of different ads. This is because the advertiser won two independent auctions, which belong to the same webpage. Moreover, when two ads share the same topic are displayed within a webpage, the probability of being competitive ranges from 21.3\% to 30.5\% in Yahoo website, and from 3.59\% to 4.55\% in MSN website. Considering the extremely large number of daily page views, the number of webpages with competitive ads is still impressive. Since user's brand conception towards displayed ad will be poor in competitive context, the effectiveness of advertising of the whole system will be affected.   

Fig.~\ref{fig:variable_distribution} shows the distribution of multimedia metrics, namely, the ad memorability, webpage-ad relevance and ad visual saliency in Yahoo and MSN. As we can see, in both two websites, the memorability score of most ads is above 0.70, indicating that these ads are well designed to be memorable by the advertisers. While contextual relevance score of the majority ads is pretty low, which is consistent with the finding from marketing data that only 2.8\% users thought ads on website were relevant~\cite{hubspot_annoy_ad}. In addition, most displayed ads are not salient, which can be easily overlooked by the user. The irrelevant and non-salient ads will result in less user engagement. In this regard, our proposed model incorporating multimedia metrics can help improve the effectiveness of existing advertising systems.

\begin{table*}[t]
\footnotesize
\centering
\caption{Summary of optimal trade-offs among stakeholders in the Yahoo (I) dataset where $\theta_1 = -0.05$. Note that $\gamma_{i}, i = 1,\dots,6$ represent the optimal weights obtained from training set, and $\xi_{i}, i = 1,\dots,6$ represent the changes of corresponding variables.}
\vspace{-5pt}
\label{tab:yahoo_10_fold_cross_validation_result}
\begin{tabular}{r|r|r|r|r|r|r|r|r|r|r|r|r|r|r|r|r|r|r}
\hline
\multirow{2}{*}{Fold}     & \multicolumn{6}{c|}{Optimal weight}	& \multicolumn{6}{c|}{Training set} & \multicolumn{6}{c}{Test set}\\
\cline{2-19}
	& $\gamma_{1}$ & $\gamma_{2}$ & $\gamma_{3}$ & $\gamma_{4}$ & $\gamma_{5}$ & $\gamma_{6}$ & $\xi_{1}$ & $\xi_{2}$ & $\xi_{3}$ & $\xi_{4}$ & $\xi_{5}$ & $\xi_{6}$ & $\xi_{1}$ & $\xi_{2}$ &
 $\xi_{3}$ & $\xi_{4}$ & $\xi_{5}$ & $\xi_{6}$\\
\hline
1 	& 0.40 & 0.35 & 0.00 & 0.05 & 0.15 & 0.05 & -3.8\% & 1.6\% & 0.0\% & 14.6\% & 0.4\% & 8.2\% & -3.5\% & 0.8\% & 0.2\% & 14.4\% & 1.0\% & 6.1\% \\ 
2	& 0.40 & 0.35 & 0.00 & 0.05 & 0.15 & 0.05 & -3.7\% & 1.4\% & 0.0\% & 14.5\% & 0.5\% & 7.8\% & -4.8\% & 2.3\% & 0.0\% & 15.0\% & 0.0\% & 9.9\% \\
3	& 0.40 & 0.35 & 0.00 & 0.05 & 0.15 & 0.05 & -3.8\% & 1.5\% & 0.0\% & 14.6\% & 0.3\% & 8.1\% & -3.9\% & 1.8\% & 0.0\% & 14.2\% & 1.6\% & 7.0\% \\
4 	& 0.40 & 0.35 & 0.00 & 0.05 & 0.15 & 0.05 & -3.9\% & 1.5\% & 0.0\% & 14.7\% & 0.7\% & 8.0\% & -3.4\% & 2.1\% & 0.3\% & 13.5\% & -2.1\% & 7.7\% \\
5	& 0.40 & 0.35 & 0.00 & 0.05 & 0.15 & 0.05 & -3.7\% & 1.3\% & 0.0\% & 14.4\% & 0.4\% & 7.9\% & -4.6\% & 3.8\% & 0.2\% & 15.8\% & 1.3\% & 9.1\% \\
6	& 0.50 & 0.40 & 0.00 & 0.00 & 0.05 & 0.05 & -1.9\% & 1.5\% & 0.0\% & 0.0\% & 0.7\% & 7.9\% & -1.7\% & 1.2\% & -0.3\% & 0.0\% & 3.4\% & 6.9\% \\
7   & 0.40 & 0.35 & 0.00 & 0.05 & 0.15 & 0.05 & -3.9\% & 1.8\% & 0.0\% & 14.5\% & 0.7\% & 7.9\% & -3.1\% & -0.2\% & 0.0\% & 15.0\% & -1.5\% & 8.6\% \\
8	& 0.40 & 0.35 & 0.00 & 0.05 & 0.15 & 0.05 & -3.8\% & 1.5\% & 0.0\% & 14.7\% & 0.5\% & 8.0\% & -4.1\% & 2.2\% & 0.4\% & 13.0\% & -0.3\% & 8.3\% \\
9	& 0.40 & 0.35 & 0.00 & 0.05 & 0.15 & 0.05 & -3.8\% & 1.5\% & 0.0\% & 14.4\% & 0.4\% & 8.2\% & -3.5\% & 2.1\% & 0.0\% & 15.7\% & 0.8\% & 6.1\% \\
10	& 0.40 & 0.35 & 0.00 & 0.05 & 0.15 & 0.05 & -3.7\% & 1.6\% & 0.0\% & 14.6\% & 0.3\% & 7.8\% & -4.3\% & 1.0\% & 0.1\% & 14.1\% & 1.8\% & 9.6\% \\
\hline
Mean	& - & - & - & - & - & - & -3.6\% & 1.5\% & 0.0\% & 13.1\% & 0.5\% & 8.0\% & -3.7\% & 1.7\% & 0.1\% & 13.1\% & 0.6\% & 7.9\% \\ 
Std.	& - & - & - & - & - & - & 0.000 & 0.000 & 0.000 & 0.001 & 0.000 & 0.000 & 0.000 & 0.000 & 0.000 & 0.001 & 0.000 & 0.000\\
\hline
\end{tabular}
\vspace{5pt}
\end{table*}

\subsection{Experimental Settings}
\label{sec:experimental_setting}

To validate the proposed framework, both bidding information (e.g., bid price, ad CTR) and multimedia information (e.g., text for contextual relevance matching, ad image for visual saliency and image memorability) for each ad are needed. As each of our datasets only provides partial information, we do random sampling to connect the auction dataset with the multimedia datasets. For each ad slot within a multi-slot webpage, we simulate the corresponding auction following the procedures described in our previous work~\cite{chen2017RTBoptimizing}: for a given webpage, the original ad (i.e., the one displayed in the webpage when we collected the data) is treated as the ground truth, and is allocated with the highest bid price. We then sample the rest candidate ads and randomly match them with bid prices. It should be noted that slots have different shapes and the ads with similar shapes can be selected as candidate ads. We then apply the framework in Section~\ref{sec:trade_offs_optimization} to select the proper combination of ads. 

Every millisecond matters in RTB. Using the TF-IDF technique~\cite{MacKay2003} to measure the contextual relevance has been demonstrated to be efficient; using MemNet~\cite{khosla2015understanding} to obtain the ad image memorability score can be conducted off-line; however, using MBS~\cite{zhang2015minimum} to calculate the visual saliency can be relatively time consuming. The processing speed of MBS is 80 frames per second on a machine with 3.2GHz$\times$2CPU and 12GB RAM. Given a webpage with 2 ad-slot, and both the two slots have 10 bidders. According to our framework, we need to calculate the saliency map for 100 images. Thus, the time consumption becomes intolerable. In our experiment, we find that: in terms of saliency, the interaction among the ad slots is small. This is because the relative locations of two ad slots is relatively far and the saliency of each slot is mainly affected by the surroundings. In this regard, we only have to calculate the saliency for 20 images in the previous example. We can further sacrifice the performance in speeding up the saliency calculating process, such as using super-pixel segmentation~\cite{xiang2015salad}.

\subsection{Results}
We conduct 10-fold cross validation on all the four datasets to demonstrate the effectiveness of our proposed framework. We first obtain the optimal weights from the training set using the algorithm described in section ~\ref{sec:optimal_weights}, then apply the optimal weights to select the most suitable ads in the test set using the optimization model described in section ~\ref{sec:trade_offs_optimization}. Note that the optimal weights change with the threshold value $\theta_{k},k=1\dots6$, which indicate publisher's preferences towards the corresponding variables. As a monetization industry, revenue is always the primary concern. For simplicity but without lose of generality, we set $\theta_{1}$ to be a non-positive value and $\theta_{k} = 0,k=2\dots6$ in our experiment. So that we will figure it out that: under the premise of the advertiser's satisfaction and the user's online experience, how much $\theta_{1}$ will change the performance of the whole system as well as the performance of other variables.  

Fig.~\ref{fig:sum_rank_score_with_theta} shows the effect of $\theta_{1}$ on the performance of the advertising system. As can be seen, in both training set and test set under all datasets, when $\theta_{1}$ decreases, the sum of rank scores of the selected ads will not change at the very beginning, decrease slightly afterwards, increase gradually later and remain stable finally. The convexity changing pattern of the sum of rank scores is such because: when $\theta_{1}$ is small, our weights determination model is not able to find a solution to the problem. In other words, the publishers are too stingy to lose revenue, and too greedy to achieve improvement on the other variables. In this case, we will use the traditional auction mechanism to select the ad with highest bidding price in each auction. When $\theta_{1}$ further decreases, the solution space increases and we will select the solution with the highest sum of rank scores. Note that, the ground truth may not be a solution to our model since competitive ads may occur. When $\theta_{1}$ decreases below a threshold value, the optimal solution will not change, which results in the stability of the sum of rank scores. It should be noted that within the same dataset, the sum of rank scores in the training set is roughly 9 times that of the test set, since the size of training set is 9 times of the size of test set in a 10-fold cross validation. 

Fig.~\ref{fig:variable_change_with_theta} describes the effect of $\theta_{1}$ on the change of variables. When $\theta_{1}$ is close to 0, all the variables remains unchanged, which is consistent with the results in Fig.~\ref{fig:sum_rank_score_with_theta}. When $\theta_{1}$ further decreases, the revenue follows a monotone decreasing pattern and the other variables increase to various degrees. We observe that when publishers specify the value of $\theta_{1}$, our optimal weights may result in much less revenue loss. For example, in Yahoo I dataset, when we set $\theta_{1}$ as -0.15, the loss of revenue obtained by our model is -0.04. This is because our framework focuses on the performance of the whole system rather than approximate the value of $\theta_{1}$. The loss of revenue acts as one of the constraints. Table 5 presents a more detailed result of a 10-fold cross validation when we set $\theta_{1}$ as -0.05 in Yahoo (I) dataset. The average loss of revenue is 3.6\%, while the average improvement of ad CTR is 13.1\% and average improvement of visual saliency is 8.0\%.  The performance of our proposed framework is very similar in all the folds, except fold 6. These consistent results confirm our cross validation. We observe that, in table~\ref{tab:yahoo_10_fold_cross_validation_result}, the weigh of memorability is 0.00, thus the change of memorability is quite small. 
This is because the optimal weights $\gamma_{k},k=1\dots6$ change with threshold values $\theta_{k}, k=1\dots6$. If the publisher values more about the memorability and assigns $\theta_{3}$ to a relatively larger value, the corresponding weight and variable change will increase.


\section{Conclusion}
\label{sec:conclusion}

In this paper, we discuss a computational framework that brings multimedia metrics to RTB to optimize trade-offs among stakeholders in display advertising. We consider the contextual relevance to ensure the user's online experience and increase CTR. We consider the visual saliency and image memorability to increase the user's engagement towards displayed ads. Our experimental results show that the proposed framework is able to increase the benefits of selected advertisers and the user with just a slight decrease in the publisher's revenue. In the long run, better engagements of advertisers and users will increase the demand of advertising and supply of webpage visits, which will boost the publisher's revenue. How to effectively model the changes and engagement process in the online advertising ecosystem in a long term will be our future work. And also, analyzing the properties of the proposed framework from the aspect of game theory will be another interesting topic. 

\section{Acknowledgement}
This research is supported by the National Research Foundation, Prime Minister's Office, Singapore under its International Research Centre in Singapore Funding Initiative.
The authors also gratefully acknowledge the support of NVIDIA Corporation with the donation of the Titan Xp GPU used for this research.

\bibliographystyle{abbrv}
\bibliography{mybib}

\end{document}